\documentclass[12pt,preprint]{aastex}

\slugcomment{Not to appear in Nonlearned J., 45.}
\shorttitle{Early optical afterglow of GRB030329} 
\shortauthors{Urata et al.} 
\begin{document}
\title{Early ($<$0.3 day) R-band light curve of the optical afterglow of GRB030329}

\author{
Yuji \textsc{Urata}\altaffilmark{1,2},
Takashi \textsc{Miyata}\altaffilmark{3},
Shingo \textsc{Nishiura}\altaffilmark{3},
Toru \textsc{Tamagawa}\altaffilmark{1},
R.A. \textsc{Burenin}\altaffilmark{4}
Tomohiko \textsc{Sekiguchi}\altaffilmark{5},
Seidai \textsc{Miyasaka}\altaffilmark{6},
Chiaki \textsc{Yoshizumi}\altaffilmark{7},
Junzi \textsc{Suzuki}\altaffilmark{8}\\
Hiroyuki \textsc{Mito}\altaffilmark{3},
Yoshikazu \textsc{Nakada}\altaffilmark{3},
Tsutomu \textsc{Aoki}\altaffilmark{3}, 
Takao \textsc{Soyano}\altaffilmark{3},
Kenichi \textsc{Tarusawa}\altaffilmark{3},
Shigetomo \textsc{Shiki}\altaffilmark{1},
and 
Kazuo \textsc{Makishima}\altaffilmark{1,9}
}
\altaffiltext{1}{RIKEN (Institute of Physical and Chemical Research), 2-1 Hirosawa, Wako, Saitama 351-0198, Japan}
\email{urata@crab.riken.go.jp}
\altaffiltext{2}{Department of Physics, Tokyo Institute of Technology, 2-12-1 Oookayama, Meguro-ku, Tokyo 152-8551, Japan}
\altaffiltext{3}{Kiso Observatory, Institute of Astronomy, The University of Tokyo, Mitake-mura, Kiso-gun, Nagano 397-0101, Japan}
\altaffiltext{4}{Space Research Institute, Moscow, Russia}
\altaffiltext{5}{National Astronomical Observatory, Osawa, Mitaka, Tokyo 181-8588, Japan}
\altaffiltext{6}{Tokyo Metropolitan Governments, Nishi-Shinjyuku 2-8-1, Shinjyuku-ku, Tokyo 163-8001 Japan}
\altaffiltext{7}{Tokushima  Science Museum, 45-22  Kibigatani,Nato,Itano-cho,Itano-gun,Tokushima 779-0111,Japan}
\altaffiltext{8}{Department of Physics, The Tokyo University of Science, 1-3 Kagurazaka, Shinjyuku, Tokyo 162-8601, Japan}
\altaffiltext{9}{Department of Physics, The University of Tokyo, 7-3-1 Hongo, Bunkyo-ku, Tokyo 113-0033, Japan}

\begin{abstract}
We observed the optical afterglow of 
the bright gamma-ray burst GRB030329 
on the nights of 2003 March 29, using the Kiso observatory
(the University of Tokyo) 1.05 m Schmidt telescope.  Data were taken
from March 29 13:21:26 UT to 17:43:16 (0.072 to 0.253 days after the
burst), using an $Rc$-band filter.  The obtained $Rc$-band light curve
has been fitted successfully by a single power law function with decay
index of $0.891\pm0.004$.  These results remain unchanged when
incorporating two early photometric data points at 0.065 and 0.073
days, reported by Price et al.(2003) using the SSO 40 inch telescope,
and further including RTT150 data (Burenin et al. 2003) covering at
about 0.3 days.  Over the period of 0.065-0.285 days after the burst,
any deviation from the power-law decay is smaller than $\pm$0.007 mag. 
The temporal structure reported by Uemura et al. (2003) does not show
up in our $R$-band light curve.

\end{abstract}

\keywords{Gamma-ray Bursts: afterglow}

\section{Introduction}
The bright and long gamma-ray burst (GRB), GRB030329, was detected on
2003 March 29 11:37:14.67 UT, with the {\it HETE-2} spacecraft (Ricker
et al. 2003).  The burst lasted for more than 25 s in the 30-400 keV
band. The fluence of the burst was ~1 x 10$^{-4}$ ergs cm$^{-2}$ and
the peak flux over 1.2 s was $>$ 7 $\times$ 10$^{-6}$ ergs cm$^{-2}$
s$^{-1}$ (i.e., $>$ 100 $\times$ Crab flux) in the same energy band.
The subsequent ground analysis of the {\it HETE-2} data produced an
accurate location for the burst, which was reported in a GCN Position
Notice at 12:50:24 UT, 73 minutes after the burst.  The location is
centered at $\alpha^{2000} = 10^{\rm h}44^{\rm m}49^{\rm s},
\delta^{2000} = +21^{\circ} 28' 44''$, with the 90\%-confidence error
radius of $2'$ (Vanderspek et al. 2003).

The optical afterglow was found within the $2'$-radius error circle at
0.077 days after the burst, at the coordinates of $\alpha^{2000}=
10^{\rm h}44^{\rm m}49.^{\rm s}5, \delta^{2000}=+21^{\circ}31'23''.1$
(Peterson and Price 2003; Torii 2003). The redshift of GRB030329 was
determined as 0.168 $\pm$ 0.001 by numerous absorption and emission
lines in the optical spectra of the afterglow (Greiner et al. 2003;
Price et al. 2003).  In the afterglow spectrum, a bright Type Ic
supernova feature appeared at $\sim$7.6 days; the associated supernova
has been named SN2003dh (e.g., Stanek et al.  2003; Hjorth et
al. 2003; Chornock et al. 2003; Zaritsky et al. 2003). This detection
strongly suggests a physical link between GRBs and supernovae
explosion.  Uemura et al. (2003) claim that the optical light curve of
the afterglow, produced by their unfiltered observation, deviates from
a single power-law decay, exhibiting temporal breaks at
0.085$\pm$0.028, 0.163$\pm$0.060, and 0.227$\pm$0.043 days.

We have established a GRB follow-up observation system at Kiso
observatory (Urata et al. 2003). Because the Japan area had been blank
for the GRB follow-up observational network, this observational system
is very important in studying the temporal and spectral evolution of
early ($<$1 day) afterglows.  Another merit of the system is its
multi-color capability, using two instruments (2k$\times$2k CCD and
KONIC) and $B$-, $V$-, $Rc$-, $Ic$-, $J$-, $H$-, and $K$-bandpass
filters (Urata et al. 2003).  We have already performed early optical
color measurements of a number of GRB afterglows.  In the present
burst, we have also performed an early $Rc$-band follow-up observation
of the afterglow, from 0.072 to 0.253 days. 
Here we study the temporal evolution of the optical afterglow over a
very early phase (0.065 to 0.286 days), by combining our Kiso
measurements with the published photometric points from the Siding
Spring Observatory (SSO0 40-inch telescope (Price et al.2003), and the
1.5 m Russian-Turkish Telescope (RTT150) by Burenin et al. (2003).

\section{Observations}

We carried out $Rc$-band follow-up observations of the optical
afterglow of GRB030329, using the system described in section
1. Specifically, we used the 1.05 m Schmidt telescope and a
2k$\times$2k CCD Camera at Kiso observatory, the University of
Tokyo. The field of view was $51.'2\times51.'2$ and the pixel size was
$1''.5$ square.  The observation started at 2003 March 13:21:26 UT
(0.075 days), and ended at 17:43:16 (0.253 days).  Meanwhile, we
acquired 36 $Rc$-band images in total, mostly with 60-s exposure.  An
example of the $Rc$-band image we obtained is shown in figure 1, where
the afterglow is extremely bright.

\section{Analysis}
The data reduction was carried out by a standard method using the NOAO
IRAF. We performed the dark-subtraction and flat-fielding correction
using appropriate calibration data.  The photometric calibrations were
done for all frames using APPHOT package in IRAF, with 7 standard
stars around the afterglow suggested by Henden (2003). We indicate
these standard stars in figure 1.  For each data, we set the
one-dimensional aperture size to 4 times as large as the full-width at
half maximum of the objects.  The magnitude error in each optical
images is estimated as $\sigma_{\rm e}^{2}=\sigma_{\rm ph}^{2} +
\sigma_{\rm sys}^{2}$, where $\sigma_{\rm ph}$ is the photometric
errors of GRB030329 afterglow, estimated from the output of IRAF PHOT,
and $\sigma_{\rm sys}$ is the photometric calibration error estimated
by comparing our instrumental magnitude of the 7 standard stars over
the 36 frames. The typical errors are $0.001\sim0.003$
mag. for photometric, and $\sim0.007$ mag. for systematic.


\section{Result}
In figure 2, we plot the $Rc$-band light curve of the afterglow of
GRB030329 based on our photometry. The light curve is well fitted with
a single power law function of the form $\propto t^{-\alpha}$, where t
is the time after the burst and $\alpha$ is the decay index; we have
obtained $\alpha=0.891\pm0.004$ with reduced chi-squared
$(\chi^2/\nu)$ of 0.819 for $\nu=$ 34.  In order to better constrain
the early-time ($<$ 0.25 day) behavior of the light curve, we combined
our data with the two $Rc$-band photometric points reported by Price
et al. (2003) using the SSO 40-inch telescope; $R$=12.6$\pm$0.015 mag. 
at 0.065 day, and $R$=12.786$\pm$0.017 mag. at 0.073 day. These are
among the earliest filtered observations of this afterglow.  We have
successfully fitted the combined $Rc$-band light curve again with a
single power law, of which the decay index is 0.891$\pm$0.003 with
$\chi^{2}/\nu$=0.817 for $\nu=36$.  We have further included 13
$Rc$-band photometric data points obtained at about 0.3 days after the
burst by Burenin et al. (2003) using the Russian-Turkish telescope
(RTT150), but the results did not change either; the decay index is
$0.890 \pm 0.03$, with $\chi^2/\nu=1.03$ for $\nu=49$.


\section{Discussion}

We observed the optical afterglow associated with GRB030329 from 0.072
to 0.253 days after the burst.  The light curve based on our
photometry is well fitted by a single power law function with a decay
index of $0.891 \pm 0.004$.  The result remains essentially unchanged
when we include the published two SSO data points (Price et al. 2003)
covering an even earlier phase (up to 0.065 days), and the published
13 RTT150 data points (Burenin et al. 2003) taken at about 0.3 days.
Thus, our single power-law representation of the $Rc$-band decay light
curve is valid altogether over a period of 0.065--0.285 days after the
burst.


Uemura et al. (2003) reported that their unfiltered light curve has
three temporal breaks, at $0.085 \pm 0.028$, $0.163 \pm 0.060$, and
$0.227 \pm 0.043$ days, across which the power-law decay index changes
by 0.2--0.5.  In figure 2, we have plotted their best-fit model
function, together with the $\pm 0.04$ magnitude constant error
suggested by them.  Because the model normalization is not specified
by Uemura et al (2003), we estimated it using 0.2 day data points of
their figure 1.  Although their measurements are thus consistent with
ours within their measurement errors, the reality of the suggested
breaks remain an important issue to be clarified.

For a more quantitative examination of the issue,
we split our data into three phases, namely
$0.065 < t < 0.163$,
$0.085 < t < 0.227$, and
$0.163 < t < 0.285$,
each covering one of the three reported breaks.  We then fitted each
of these light curve segments with a single power-law, and obtained
the results as summarized in Table 1.  Thus, the three segments are
all expressed adequately with a single power-law, and the derived
decay indices do not differ by more than 0.13.

We also tried to fit our light curve with the smoothly broken
power-law model used by Uemura et al. (2003), which is expressed as
$f(t)=\{ f_{1}(t)^{-n}+f_{2}(t)^{-n} \}^{-1/n}$ with
$f_{i}=k_{i}t^{-\alpha_ i}$ ($i=1,2$), where $t$ is the time after the
burst onset, $f(t)$ is the $R_c$-band flux, $\alpha_1$ and $\alpha_2$
are the decay indices at early and late times respectively, $k_i$ is
a normalization constant, and $n$ indicates a smoothness parameter.
The break times were fixed to the values reported by Uemura et
al. (2003).  Furthermore, we fixed $n$ at unity after Uemura et al
(2003) when fitting the $0.085 < t <0.227$ phase.  The results from
this analysis are again summarized in Table 1.  In the first and the
third time segments, the slope change is thus insignificant (at most
0.1) and the decay indices before and after the assumed break are all
consistent with the global value of 0.89.  Although a break is
apparently suggested by the second segment, an F-test indicates that
the improvement of the broken power-law model over the single
power-law is not significant at the 90\% confidence level.  Thus, our
high-quality $Rc$-band light curve does not show any evidence for the
wriggle structure reported by Uemura et al. (2003).

We further tried to fit the present light curve (Kiso, SSO, and
RTT150) with the best-fit model function reported by Uemura et
al. (2003), in which the over all normalization alone is allowed to
vary.  The fitting has yielded with $\chi^2/\nu=6.91$ for $\nu=50$.
Based on this large value of $\chi^2/\nu$, we can rule out the model
of Uemura et al. (2003) at more than 99.99\% confidence level.
%
%
A discrepancy between the filtered and unfiltered light curves could
arise if there were significant color changes during the decay.  In
this case, a similar deviation from a single power-law decay should be
observed in unfiltered light curves by other observers.  However,
there is no such structures in the unfiltered light curve obtained by
Torii et al. (2003); it is fitted successfully by a single power law
with a decay index of $0.891 \pm 0.016$, in a good agreement with the
present result.
Furthermore Uemura et al. (2003) report that their unfiltered CCD has
response close to that of the $Rc$- system. Therefore, the origin of
the discrepancy must be found somewhere else.
%

\acknowledgments 
We thank many high school students who participated in the education
program at Kiso observatory in the night to collaborate with us. We
also thank RTT150 collaboration for their observations as well as for
making their data publicly available.  Y. Urata acknowledges support
from the Japan Society for the Promotion of Science (JSPS) through
JSPS Research Fellowships for Young Scientists.


\begin{table}
\begin{center}
\caption{Results of fitting the present $Rc$-band light curve.}
\begin{tabular}{lcccccc}
\hline
 Interval      & $\alpha ^a$   &break$ ^b$&$\alpha_1~^b$    &$\alpha_2~^b$ & $\chi^2/\nu$ & $\nu$  \\
\hline
$0.065<t<0.163$&$0.900 \pm0.011$ &  --- &   ---             &  --- & 0.58       & 12 \\
               &    ---          &0.085d& $0.906 \pm 0.001$ & $0.895 \pm 0.001$ & 0.69$^c$ & 11 \\
$0.085<t<0.227$&$0.897 \pm0.005$ &  --- &   ---             &  --- & 0.80       & 27 \\
               &    ---          &0.163d& $0.949 \pm 0.050$ & $0.844 \pm 0.053$ & 0.49     & 25 \\
$0.163<t<0.285$&$0.877 \pm0.007$ &  --- &   ---             &  --- & 1.10       & 35 \\
               &    ---          &0.227d& $0.881 \pm 0.001$ & $0.873 \pm 0.001$ & 1.20     & 33 \\
\hline
\end{tabular}
\end{center}
$^a$ A single power-law fit, with $\alpha$ being the decay index.\\
$^b$ A fit with the smoothly broken power-law model described in the text.\\
$^c$ The smoothness parameter is fixed at $n=1$.
\end{table}

\begin{figure} 
\plotone{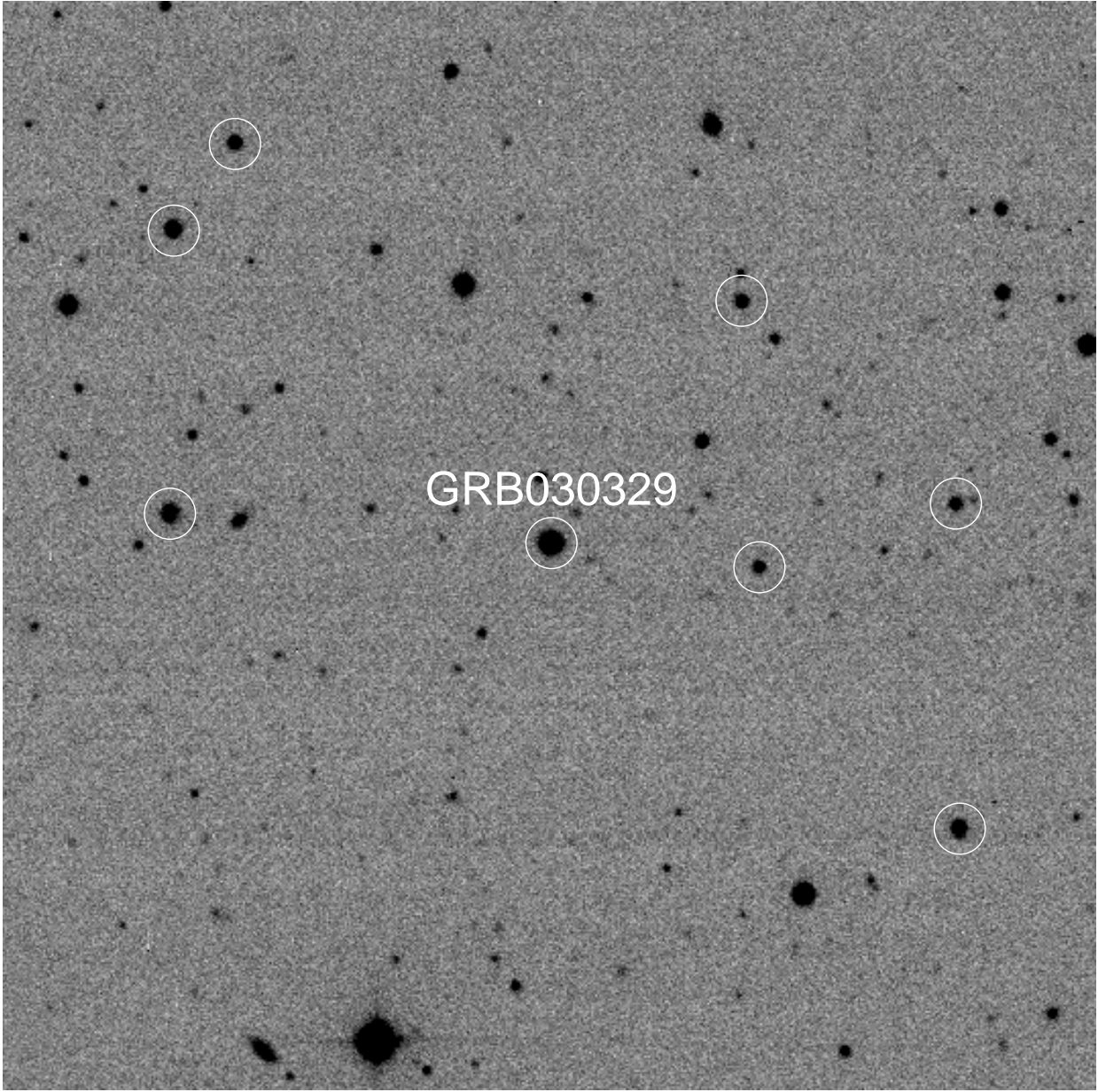} 
\caption{An $Rc$-band image of the GRB030329 field obtained at the
 Kiso observatory, with a 60-s exposure starting at 2003 March 29
 14:23:50 UT (0.116 days after the burst). The afterglow is indicated
 by an arrow near the image center. Circles indicate the standard
 stars used in the photometric calibration.
\label{fig1}}
\end{figure}

\begin{figure}
\plotone{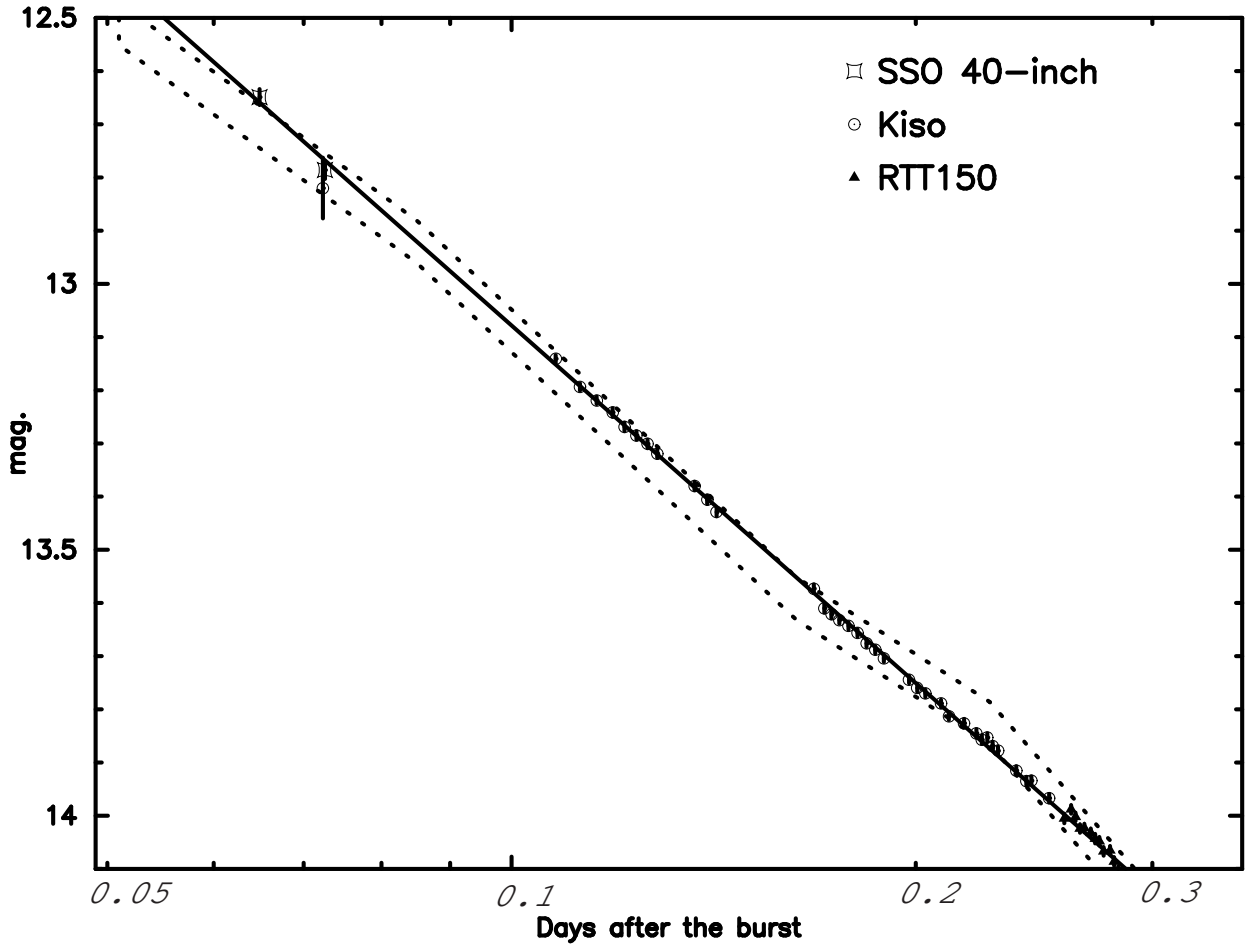}
\caption{The $Rc$-band light curve based on the photometry at Kiso,
shown together with the SSO (Price et al. 2003) and the RTT150
(Burenin et al. 2003) data points. The solid line indicates the best
fit power-law to the Kiso and SSO points. Dashed lines indicate
$\pm$0.04 mag error band around the unfiltered light curve reported by
Uemura et al (2003).}
\end{figure}

\end{document}